\documentclass[conference, compsocconf]{IEEEtran}
\IEEEoverridecommandlockouts
\ifCLASSOPTIONcompsoc
  \usepackage[nocompress]{cite}
\else
  \usepackage{cite}
\fi

\usepackage{amsmath,amssymb,amsfonts}
\usepackage{algorithmic}
\usepackage{algorithm}
\usepackage{graphicx}
\usepackage{textcomp}
\usepackage{xcolor}
\usepackage{url}
\usepackage{booktabs}
\usepackage{tablefootnote}
\usepackage{listings}
\usepackage{comment}
\usepackage{graphicx}
\usepackage{booktabs}
\usepackage{listings}
\usepackage{xcolor}
\usepackage{tablefootnote}
\usepackage{here}
\makeatletter 
\newcommand{\linebreakand}{%
  \end{@IEEEauthorhalign}
  \hfill\mbox{}\par
  \mbox{}\hfill\begin{@IEEEauthorhalign}
}

\def\BibTeX{{\rm B\kern-.05em{\sc i\kern-.025em b}\kern-.08em
    T\kern-.1667em\lower.7ex\hbox{E}\kern-.125emX}}

\begin{document}

\title{3Dify: a Framework for Procedural 3D-CG Generation Assisted by LLMs Using MCP and RAG}

\author{\IEEEauthorblockN{Shun-ichiro Hayashi}
\IEEEauthorblockA{Graduate School of Informatics\\
Nagoya University\\
Nagoya, Aichi 464-8601\\
Email: hayashi@hpc.itc.nagoya-u.ac.jp}
\and
\IEEEauthorblockN{Daichi Mukunoki}
\IEEEauthorblockA{Information Technology Center\\
Nagoya University\\
Nagoya, Aichi 464-8601\\
Email: mukunoki@cc.nagoya-u.ac.jp}
\and
\IEEEauthorblockN{Tetsuya Hoshino}
\IEEEauthorblockA{Information Technology Center\\
Nagoya University\\
Nagoya, Aichi 464-8601\\
Email: hoshino@cc.nagoya-u.ac.jp}
\linebreakand
\IEEEauthorblockN{Satoshi Ohshima}
\IEEEauthorblockA{
Research Institute for Information Technology\\
Kyushu University\\
Fukuoka 819-0395, Japan\\
Email: ohshima@cc.kyushu-u.ac.jp}
\and
\IEEEauthorblockN{Takahiro Katagiri}
\IEEEauthorblockA{Information Technology Center\\
Nagoya University\\
Nagoya, Aichi 464-8601\\
Email: katagiri@cc.nagoya-u.ac.jp}
}

\maketitle

\begin{abstract}
This paper proposes “3Dify,” a procedural 3D computer graphics (3D-CG) generation framework utilizing Large Language Models (LLMs). The framework enables users to generate 3D-CG content solely through natural language instructions. 3Dify is built upon Dify, an open-source platform for AI application development, and incorporates several state-of-the-art LLM-related technologies such as the Model Context Protocol (MCP) and Retrieval-Augmented Generation (RAG). For 3D-CG generation support, 3Dify automates the operation of various Digital Content Creation (DCC) tools via MCP. When DCC tools do not support MCP-based interaction, the framework employs the Computer-Using Agent (CUA) method to automate Graphical User Interface (GUI) operations. Moreover, to enhance image generation quality, 3Dify allows users to provide feedback by selecting preferred images from multiple candidates. The LLM then learns variable patterns from these selections and applies them to subsequent generations.
Furthermore, 3Dify supports the integration of locally deployed LLMs, enabling users to utilize custom-developed models and to reduce both time and monetary costs associated with external API calls by leveraging their own computational resources.
\end{abstract}


\section{Introduction}
\label{sec:introduction}
3D computer graphics (3D-CG) has become an indispensable technology for sharing three-dimensional visual information across a wide range of social domains. Its applications extend beyond entertainment industries such as movies and games to areas including product design in manufacturing, surgical simulation in healthcare, education, and digital-twin technologies that replicate the real world within virtual spaces. 3D-CG enables observation from arbitrary viewpoints, physical simulations, and interactive operations with spatial consistency. These capabilities have established it as a powerful tool for real-world problem-solving.
\par 

With the growing demand for 3D-CG production, creators are increasingly required to efficiently generate complex, large-scale scenes and diverse content variations. Procedural methods, which systematically generate content based on algorithms and rules, have emerged as a promising approach to meet this need. For example, consider the task of creating numerous houses with distinct designs in an open-world game set within a vast environment. Traditional manual modeling demands extensive effort to adjust component dimensions (e.g., wall and roof heights) and to reassemble them for each house. In contrast, procedural methods can automatically adjust such parameters by defining dependency relationships -- such as between a roof and its supporting walls -- in the form of a node graph.
Major Digital Content Creation (DCC) tools, including Houdini\footnote{https://www.sidefx.com/ja/products/houdini/} and Blender\footnote{https://www.blender.org/}, provide functionality to visually construct generation procedures through node-based graph connections. This design significantly lowers the barrier for non-expert users to adopt procedural modeling techniques.
\par 

However, several challenges remain in mastering procedural methods. First, formulating diverse patterns of target objects into generation rules and implementing them within large and complex node graphs requires advanced expertise and extensive experience. Second, the implementation of procedural generation functions varies among DCC tools, making it difficult to reuse acquired knowledge and created assets.
\par 

To address these issues, a growing body of research has explored procedural 3D-CG generation from natural language instructions using Large Language Models (LLMs). However, existing LLM-based frameworks still face many unresolved challenges. AI technology evolves at an extremely rapid pace, demanding continuous and agile adaptation to the latest models and methods. Moreover, implementations tied to specific DCC tools lack versatility and cannot be easily extended to other platforms. Furthermore, mechanisms that enable LLMs to accurately understand users’ intended designs and variations -- and to interactively refine output quality -- are still under development. In addition, enterprises often require local processing environments to avoid sending sensitive data to external API services for security reasons.
\par 

In this study, we propose ``3Dify,'' a multifunctional procedural 3D-CG generation framework powered by LLMs. In 3Dify, users can generate 3D-CG simply by providing natural language instructions to an LLM. The LLM does not directly generate 3D content; instead, it leverages the capabilities of existing Digital Content Creation (DCC) tools such as Blender, Unreal Engine\footnote{https://www.unrealengine.com} (Epic Games), and Unity\footnote{https://unity.com/ja} (Unity Technologies) to produce 3D scenes. This approach allows users to take full advantage of each tool's extensive features and high functionality. Furthermore, 3Dify introduces distinctive features such as feedback loops that help users concretize the images they wish to generate, and automated control of DCC tools. As a result, 3Dify achieves high efficiency and flexibility in procedural 3D-CG generation.
\par 

The structure of this paper is as follows. Section \ref{sec:relatedwork} introduces related work. Section \ref{sec:overview} introduces the workflow and its key features of 3Dify. Section \ref{sec:method} presents the implementation. Section \ref{sec:demo} demonstrates an example of automatic 3D model generation using only MCP. Finally, Section \ref{sec:conclusion} presents the conclusion.
\par 

\section{Related Work}
\label{sec:relatedwork}
Most Large Language Models (LLMs) are built upon neural networks employing the Transformer architecture~\cite{vaswani2017attention}. Owing to its high versatility, the Transformer has been applied not only to natural language generation but also to multimodal generation, including image synthesis. Image generation has since continued to evolve through integration with various AI technologies, such as Variational Autoencoders (VAE)~\cite{kingma2022autoencodingvariationalbayes}, U-Net~\cite{ronneberger2015unet}, and diffusion models~\cite{10.5555/3495724.3496298}.
\par 

Unlike 2D computer graphics (2D-CG), 3D-CG involves a wide range of components, including vertices, edges, faces, normals, textures, lighting, UV maps, bones, physics simulation weights, morph targets, and animations. Multiple modeling methods also exist, such as polygonal, subdivision, sculpting, and procedural modeling. Consequently, research on AI-assisted 3D-CG generation has diversified in both its objectives and methodologies.
\par 

With improvements in LLM performance, researchers have begun exploring their use in procedural 3D-CG generation. For example, Infinigen~\cite{rebain2024infinigen} is a procedural 3D-CG generation tool based on mathematical rules. 3D-GPT~\cite{three_d_gpt} sought to achieve interactive 3D-CG generation in Blender via natural language by controlling Infinigen itself through LLMs. It supports the discovery of useful parameters for procedural generation by allowing the LLM to describe details of the generated object's shape. SceneX~\cite{DBLP:journals/corr/abs-2403-15698} extended this line of research, while LL3M~\cite{lu2025ll3m} represents the latest advancement. LLMR~\cite{fernanda2024llmr} utilizes LLMs to leverage Unity's multi-platform capabilities, including features that respond to device inputs such as VR and animation generation. However, it relies on the proprietary Roslyn C\# compiler, which imposes strict Unity version constraints. Ludas AI\footnote{https://ludusengine.com} is an AI plugin for Unreal Engine; however, its LLM agents are black-boxed, and the scope of automated operations remains limited. VLMaterial~\cite{li2025vlmaterial} focuses on material generation, which governs surface appearance in procedural workflows. In addition, several studies have proposed methods for selecting and placing existing assets in 3D scenes. For example, SceneCraft~\cite{10.5555/3692070.3692846} employs CLIP, which computes similarities between images and text, to retrieve assets that match input text from asset libraries. It then utilizes a multimodal LLM's visual understanding to adjust placement within Blender. GraphDreamer~\cite{Gao_2024_CVPR} performs relational placement by having the LLM output relationship graphs among multiple objects. Furthermore, DIScene~\cite{10.1145/3680528.3687589} demonstrated superior performance in this field.
\par 

\begin{figure}[t]
    \centering
    \includegraphics[width=\hsize]{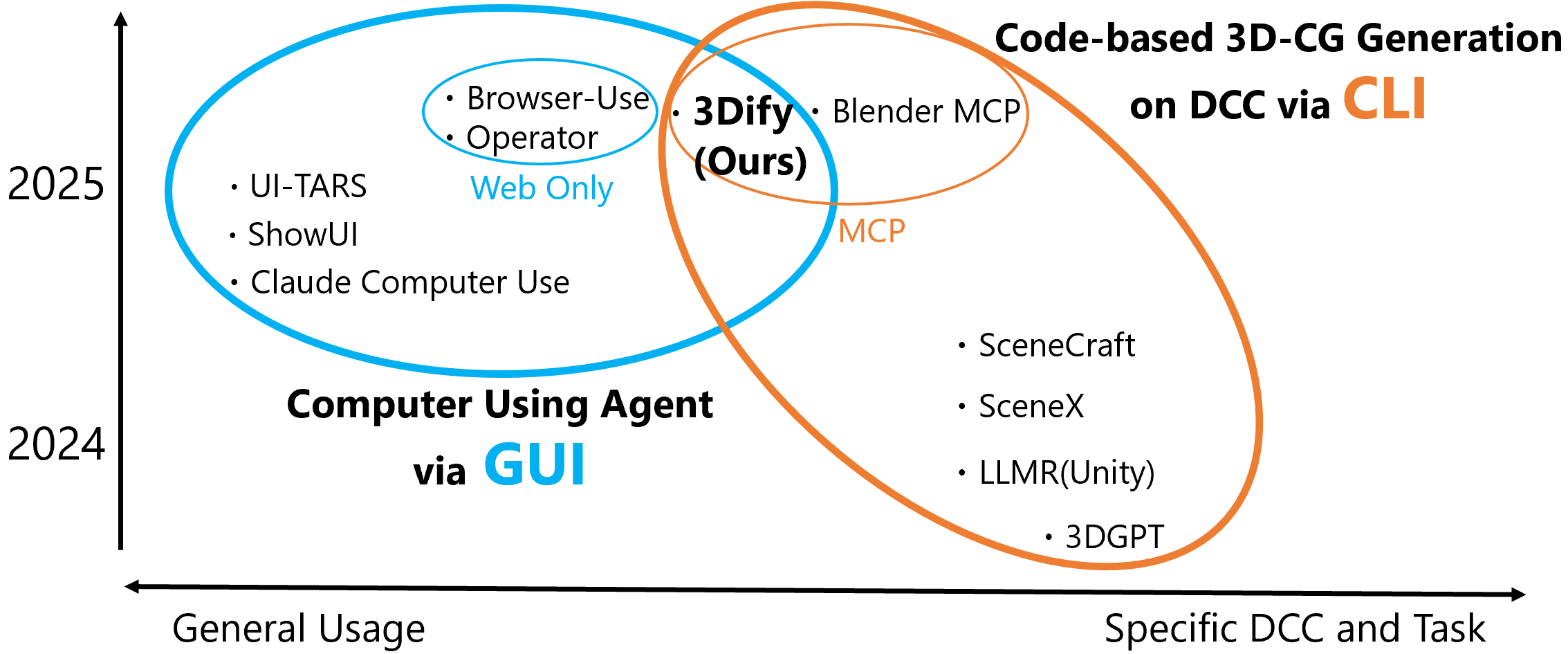}
    \caption{Position of 3Dify in related software.}
    \label{fig:classification}
\end{figure}

Fig.~\ref{fig:classification} illustrates the positioning of 3Dify relative to existing research.
\par 

\section{Overview of 3Dify}
\label{sec:overview}
This section first outlines the 3Dify workflow. It then describes the features of 3Dify.
\par 

\subsection{Workflow}
The workflow for creating 3D computer graphics using 3Dify is as follows.
\begin{itemize}
    \item \textbf{Step 1. Prompt Input}: The user provides natural language instructions describing the 3D image they wish to generate.
    \item \textbf{Step 2. Feedback Loop for Concretizing the Desired Image}: Based on Step~1, an LLM presents multiple 2D (not 3D) image candidates as pre-visualization (pre-viz) images. The user then selects several images that are closest to the intended result, and the LLM generates new image candidates based on those selections. This process is iteratively repeated until an image that closely matches the user's intent is obtained. Details of this process are described in Section~\ref{sec:feedback}.
    \item \textbf{Step 3. 3D Image Creation via Automated DCC Tool Operation}: Based on the outcome of Step~2, a DCC tool automatically creates the corresponding 3D image. The operation of the DCC tool is automated. Details of this process are described in Section~\ref{sec:autooperation}.
\end{itemize}
A key point to note is that, in the ideal scenario, users can generate 3D images solely through natural language instructions and image selection tasks, without manually performing any 3D modeling operations within DCC tools. The next section explains how this is achieved.
\par 

\subsection{Key Features}
3Dify realizes the following features:

\begin{itemize}
\item \textbf{Dify-based implementation}: 3Dify is built by extending Dify\footnote{https://dify.ai}, an open-source platform for AI application development provided by LangGenius. Because Dify rapidly integrates a wide range of state-of-the-art AI technologies, 3Dify can immediately take advantage of them. For example, it allows easy switching among the latest LLM models provided by OpenAI\footnote{https://openai.com}, Anthropic\footnote{https://docs.anthropic.com}, Google\footnote{https://gemini.google.com}, and others. Furthermore, being open source ensures platform continuity, functional extensibility, and long-term maintainability in the future.
\item \textbf{Automatic operation of DCC tools}: 3Dify provides mechanisms to automatically operate DCC tools such as Blender, Unreal Engine, and Unity via Model Context Protocol (MCP)~\cite{hou2025model} (described in Section \ref{sec:mcp}) and Computer Using-Agent (CUA) (described in Section \ref{sec:cua}).  
\item \textbf{Utilization of RAG}: Retrieval-Augmented Generation (RAG)~\cite{lewis2020retrieval} is a method for enhancing generation capability by referencing external information. 3Dify utilizes RAG to improve functionality, maintainability, and 3D-CG generation capability.  
\item \textbf{Image-selection feedback loop}: To help LLMs generate better image candidates, 3Dify provides a feedback function. Users can select multiple candidates from various generated images as feedback. This allows the LLM to automatically recognize variable patterns from these images and utilize them for subsequent image generation.  
\item \textbf{Allowing the use of local LLMs}: 3Dify enables the use of local LLMs, allowing users to utilize their own computational resources. This reduces API costs (time and fees) associated with external LLM services (such as those provided by OpenAI) and also allows the use of custom models. Furthermore, it prevents data leakage to external systems when handling sensitive information.  
\item \textbf{Extensibility beyond 3D-CG production}: 3Dify can be used not only for 3D-CG production but also for various related tasks, which is one of its major advantages over existing frameworks. The use of CUA enables access to all features supported by DCC tools. Some DCC tools provide not only 3D-CG creation functions but also a wide range of capabilities, such as game development and animation creation. Because these can be operated with a unified user interface similar to that used for 3D-CG generation, CUA operations can be directly applied. The use of RAG also supports this extensibility. Referencing documentation for each DCC tool further enhances support for operations of functions beyond 3D generation.  
\end{itemize}
\par 

\begin{figure*}[t]
\centering
\includegraphics[width=0.65\hsize]{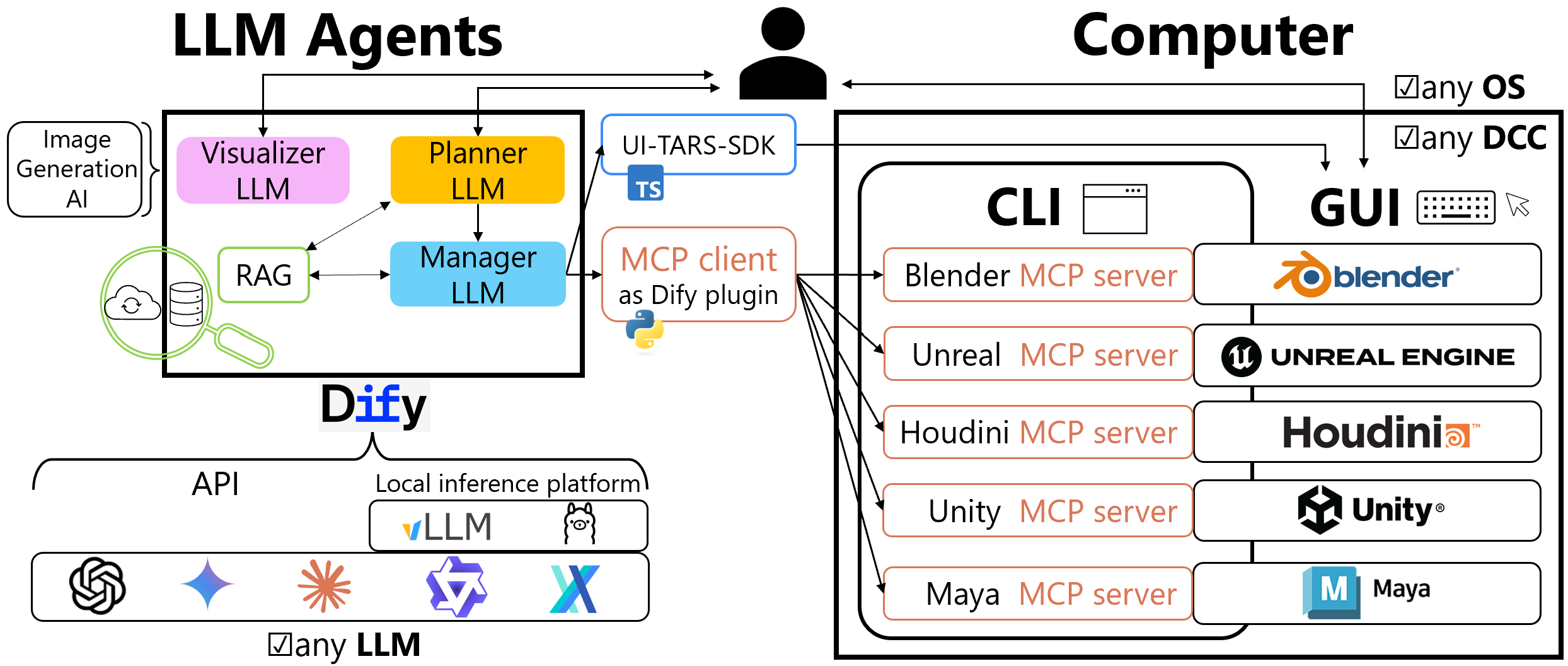}
\caption{System architecture diagram of the 3Dify framework}
\label{fig:3dify}
\end{figure*}

\begin{figure*}[t]
\centering
\includegraphics[width=0.65\hsize]{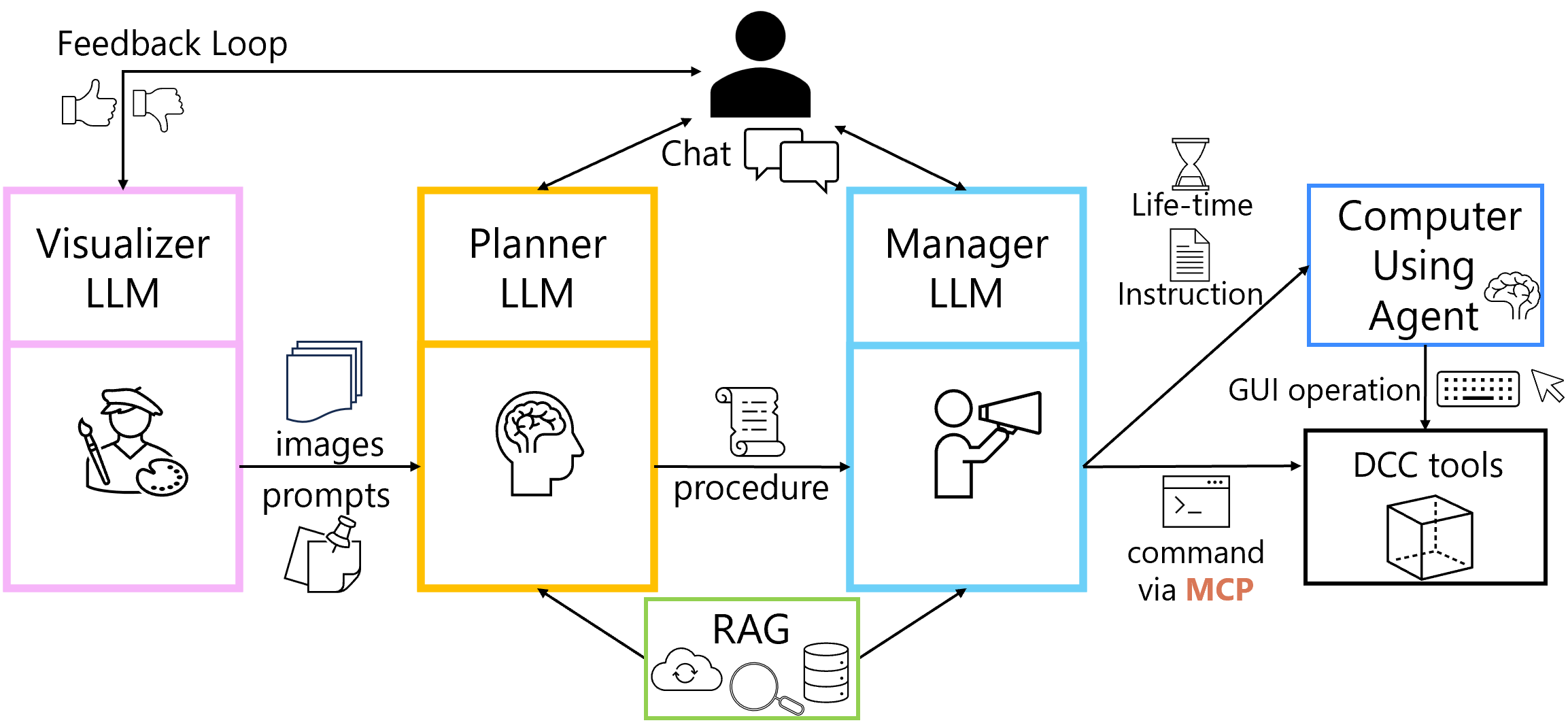}
\caption{Detailed view of 3Dify's LLM agents}
\label{fig:3dify-detail}
\end{figure*}

\section{Implementation of 3Dify}
\label{sec:method}
This section describes the implementation of 3Dify's main components. Fig.~\ref{fig:3dify} shows the overall structure of 3Dify, which consists of (1) three LLMs with distinct roles (Planner LLM, Manager LLM, and Visualizer LLM), (2) RAGs referenced by the Planner LLM and Manager LLM, (3) an iterative image-generation feedback loop, and (4) MCP servers for each DCC tool along with an MCP client for Dify. In this figure, the ``Computer'' on the right represents the machine on which the DCC tools are installed. Normally, users operate these tools via a GUI, but in 3Dify, the Manager LLM performs these operations automatically.
\par 

\subsection{Configuration of Multiple LLM Agents}
3Dify employs three LLMs with different roles as follows:

\begin{itemize}
\item \textbf{Visualizer LLM}: it generates pre-visualization (2D-CG) candidate images based on the user's natural language instructions. Then, based on user feedback, it regenerates and presents the pre-visualization images again.
\item \textbf{Planner LLM}: it receives pre-visualization images from the Visualizer LLM, predicts the required variability for the 3D model from among them, extracts the procedural model's parameters and their scope, and communicates the procedure for reflecting this into the 3D model to the Manager LLM.
\item \textbf{Manager LLM}: it receives instructions from the Planner LLM and operates the DCC tool for 3D-CG creation. It also assists this process through interaction with the user as needed.
\end{itemize}

Fig.~\ref{fig:3dify-detail} shows the correspondence between the user's workflow and the roles of each LLM.
\par 





\begin{figure}[t]
\centering
\includegraphics[width=\hsize]{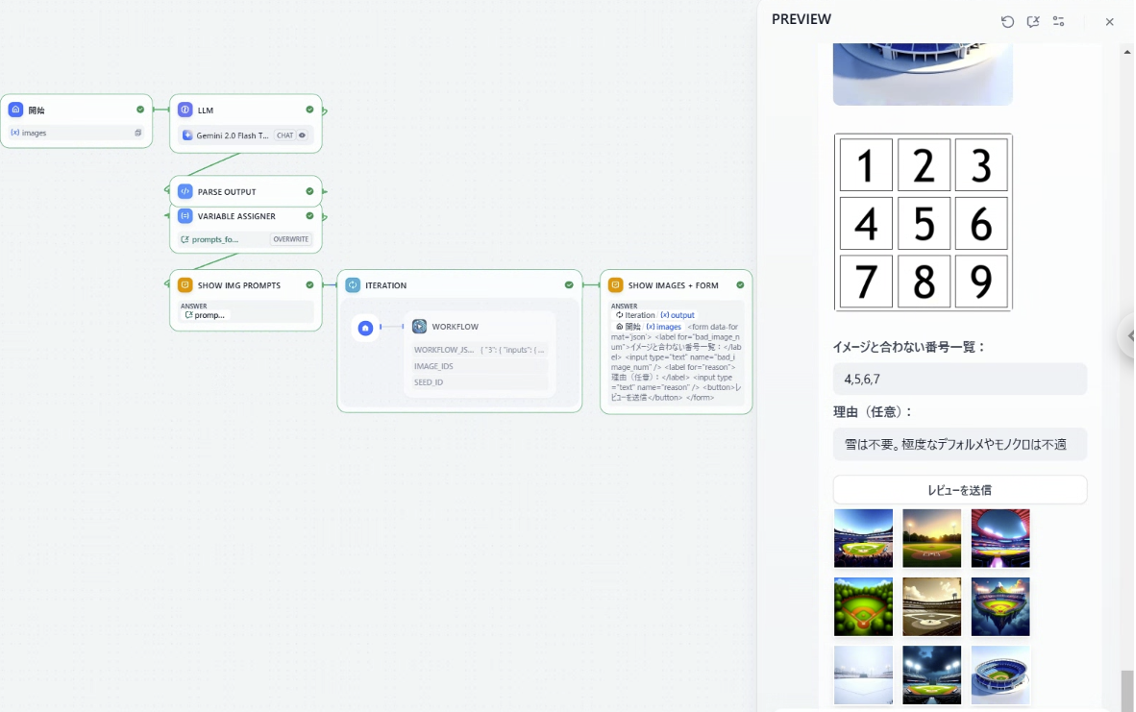}
\caption{Image-selection feedback loop.}
\label{fig:feedback}
\end{figure}

\subsection{Image-selection Feedback Loop}
\label{sec:feedback}
3Dify provides a mechanism to improve image generation quality by enabling users to create images interactively and iteratively with the system. First, the LLM generates several candidate images, and the user selects the most desirable ones to provide as feedback to the LLM. The LLM automatically recognizes variable patterns useful for procedural generation from these images and applies them in subsequent image generation. The specific process is as follows. Suppose the user intends to generate $n$ images.

\begin{enumerate}
\item The LLM generates $n$ candidate images. These images are low-quality, temporary outputs as 2D-CG (equivalent to so-called pre-visualization confirmation) created using a fast generation method. If the generated images are too similar and lack variation, the user can instruct the LLM to introduce more diversity by providing an additional prompt.
\item The user selects $m \le n$ images considered close to the desired set from the $n$ candidates and provides feedback to the LLM. At this time, the user may also give textual reasons for rejecting non-selected images, which the LLM can use as a reference for the next generation cycle.
\item This process is repeated until $m = n$.
\item For the final $n$ candidates, the LLM generates high-quality 3D-CG images that are no longer pre-visualizations.
\end{enumerate}

Fig.~\ref{fig:feedback} shows a screenshot of this process.
\par 
 
\subsection{Chatflow Templates for Feedback Loop}
3Dify creates 3D images by reflecting the user's intent through interactive processes. Therefore, multi-turn interaction is essential. For instance, when the Manager LLM encounters difficulties during an operation, it may consult either the user or the Planner LLM. In Dify, multi-turn interaction is implemented using a feature called Chatflow. However, Chatflow has a limitation: unless values are explicitly stored in \textit{Conversation Variables}, information cannot be carried over to the next turn. As a result, constructing robust multi-agent workflows from scratch is costly -- especially for pipelines such as 3Dify, which involve dynamic branching and looping among the Planner LLM, Manager LLM, and user.
\par 

To address this issue, we provide a Chatflow template that supports multi-agent workflows involving branching and looping. This template is utilized in the implementation of 3Dify itself and can also be reused to achieve similar functionality in other multi-agent systems such as LLMR, 3D-GPT, and LL3M.
\par 

\begin{figure*}[t]
\centering
\includegraphics[width=\hsize]{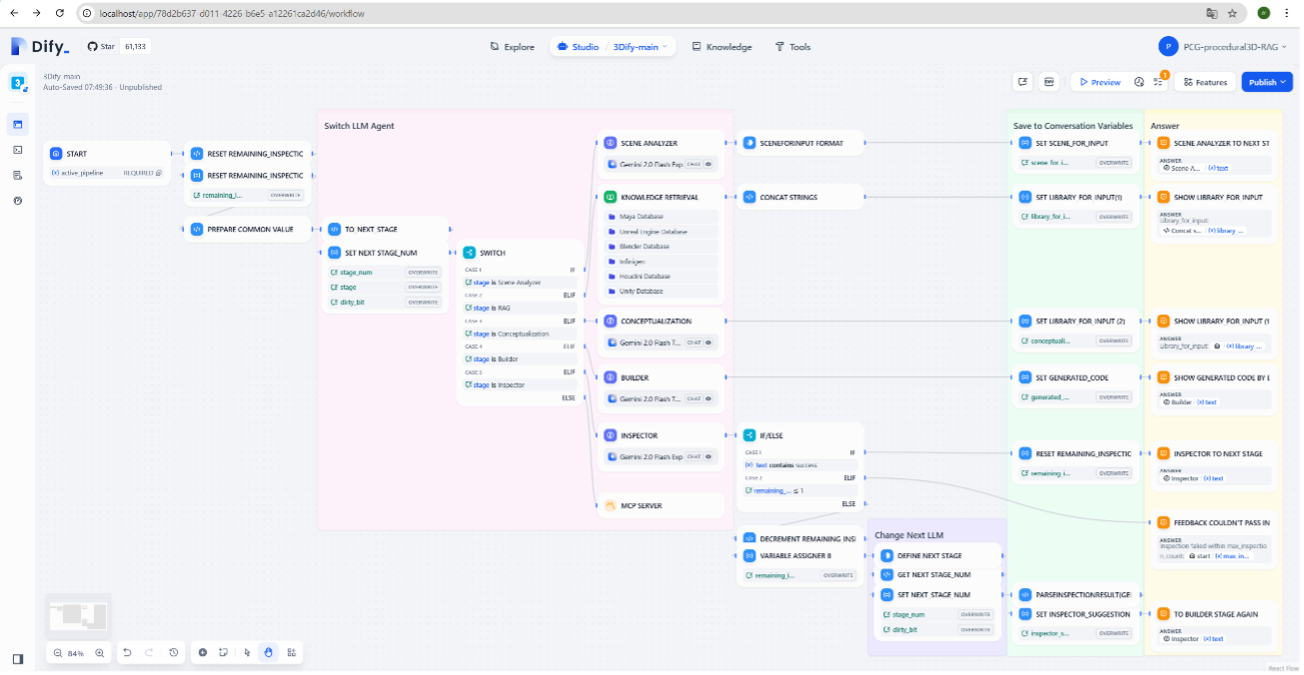}
\caption{Example of Chatflow in 3Dify}
\label{fig:9}
\end{figure*}

Fig.~\ref{fig:9} shows an example of an overall workflow implemented as a Chatflow. Processing begins from the start node on the far left. When it reaches the yellow answer node on the far right, it displays the output text and returns to user input. This process is repeated as one interaction turn. Chatflow has a specification whereby everything except conversation variables is reset when returning to the start. Therefore, any information from LLM outputs or MCP server responses that must be used in subsequent turns needs to be written to conversation variables. In the second vertical area from the right, highlighted in light green in Fig.~\ref{fig:9}, only this operation is performed.
\par 

Table~\ref{tab:agent-vars} shows several conversation variables required for multi-agents to dynamically select the next agent. The variable \texttt{stages} is a list-type conversation variable containing strings that represent the agent list. From \texttt{stages = [``Scene Analyzer'', ``RAG'', ``Conceptualization'', ``Builder'', ``Inspector'']}, the string value is extracted as \texttt{stage = stages[stage\_num]}. This value is then updated to the current stage, after which the system switches branches accordingly.
\par 

The \texttt{TO\_NEXT\_STAGE} node (the third from the left) is a function written in Python. It is designed to increment \texttt{stage\_num} and automatically proceed to the next stage when \texttt{dirty\_bit} equals 0. The \texttt{dirty\_bit} is set to 1 only when the next \texttt{stage\_num} is explicitly specified in the previous turn.
\par 

\begin{table}[t]
  \centering
  \caption{Conversation variables for multi-agent control}
  \label{tab:agent-vars}
  \begin{tabular}{lll}
    \toprule
    Variable name & Type & Description \\ \midrule
    stage                     & String & Current agent name \\ 
    dirty\_bit                & Number & Next agent decided \\ 
    enable\_increment         & Number & Auto-progress flag \\ 
    stage\_num                & Number & Agent number \\ 
    stages                    & Array[String] & Agent list \\ \bottomrule
  \end{tabular}
\end{table}

The green node labeled \textit{Knowledge Retrieval} represents RAG. Using Dify's question-classifier node, one can easily design a system that accesses separate databases for each software. However, 3Dify does not readily adopt this implementation. The reason is that we aim to share the concepts and procedures of procedural generation itself, independent of specific DCC tools. For example, when creating a room using Unreal Engine's PCG procedural generation function, we envision cases where knowledge and solvers related to indoor object placement constraints from Infinigen (originally developed for Blender) could be reused.
\par 

\begin{figure}[t]
\centering
\includegraphics[width=\hsize]{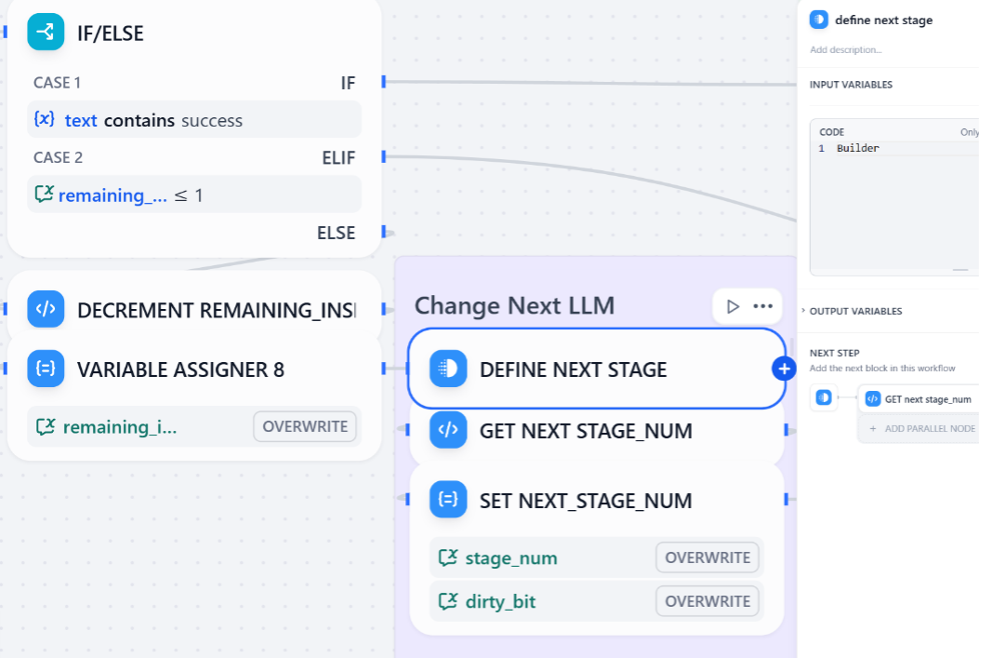}
\caption{Utility for easily implementing branching to arbitrary agents in Chatflow}
\label{fig:11}
\end{figure}

3Dify's main workflow provides a node set that facilitates easy addition or removal of stages. Even when the transition to the next stage is not a simple linear workflow such as $A\rightarrow B\rightarrow C\rightarrow ...$, it can be handled with only minor modifications. By duplicating the block enclosed in purple in Fig.~\ref{fig:11} and changing the \texttt{Builder} text in \texttt{define next stage} to the name of the next stage, the system can dynamically switch to the next agent based on the LLM's judgment.
\par 

In the internal Python code, when \texttt{SET} is performed with \texttt{dirty\_bit = 1}, the input string is converted to \texttt{stage\_num}. The \texttt{Builder–Inspector} loop is implemented by defining a conversation variable called \texttt{remaining\_inspection\_count}. This variable represents the maximum number of remaining loop iterations, as shown in Table~\ref{tab:loop-vars}, in addition to this node set. Note that the decrement operation is performed at the node located in the left-center area of Fig.~\ref{fig:11}.
\par 

\begin{table}[t]
  \centering
  \caption{Conversation variables for loop count control}
  \label{tab:loop-vars}
  \begin{tabular}{lll}
    \toprule
    Variable name & Type & Description \\
    \midrule
    max\_inspection\_count      & Number & Maximum attempts \\ 
    remaining\_inspection\_count & Number & Remaining attempts \\ 
    \bottomrule
  \end{tabular}
\end{table}

\subsection{Utilization of RAG}
\label{sec:rag}
Retrieval-Augmented Generation (RAG)~\cite{lewis2020retrieval} is a method for enhancing the generation capabilities of LLMs by referencing external information during inference. Such external information may include specially constructed databases or the Internet. 3Dify employs RAG as a key technique to improve the performance, functionality, and maintainability of 3D-CG generation.
\par 

To enhance generation performance, 3Dify leverages \textit{Parent–Child Indexing}, introduced in Dify v0.15.0. Parent–child search, a type of RAG, often utilizes embedding models optimized for vector search, such as \texttt{text-embedding-ada-002}\footnote{https://platform.openai.com/docs/models/text-embedding-ada-002}. In previous systems, similar functionality had to be implemented manually. For example, LLMR's Skill Library and 3D-GPT's Task Dispatch Agent provide lists to a conventional LLM to identify relevant child elements, then use an output parser to substitute them with different large parent files. 3Dify achieves equivalent functionality at a lower cost by using the built-in RAG features provided in Dify. This is accomplished simply by uploading related documents and selecting an appropriate chunking strategy.
\par 

For improved functionality and maintainability, RAG enables LLMs to reference manuals and documentation of DCC tools, thereby enhancing their functional coverage. It also supports flexible adaptation to functionality and operational changes arising from version upgrades of DCC tools. Because RAG can directly handle unstructured data, these capabilities can be realized easily and efficiently.
\par 

\subsection{Automatic Operation of DCC Tools}
\label{sec:autooperation}
To create 3D-CG, 3Dify utilizes existing DCC tools such as Blender, Unreal Engine, and Unity by operating them through LLMs. 3Dify provides two methods for automatic operation.
\par 

\subsubsection{Using MCP}
\label{sec:mcp}
One approach is operation via the Model Context Protocol (MCP)~\cite{hou2025model}, an open-source protocol for LLM agents released by Anthropic in 2024. MCP enables the simple and secure construction of bidirectional connections between LLM agents, applications (in this case, DCC tools), and external data sources such as RAG.
\par 

To operate DCC tools, we implemented an MCP server on each DCC tool to receive operation commands from LLM agents. Currently, 3Dify provides MCP servers for Blender\footnote{At the time of writing, another MCP server implementation has already been published: https://github.com/ahujasid/blender-mcp} and Unreal Engine only, but other DCC tools can be supported in a similar manner. To simplify the implementation of MCP servers for other DCC tools, we provide MCP server templates, as shown in Fig.~\ref{fig:mcp_template}.
\par 

On the Dify side, we implemented a prototype MCP client for communicating with MCP servers. As of this writing, although another plugin has become the official Dify MCP client, our implementation additionally provides functionality for using UI-TARS (explained in Section~\ref{sec:cua}).
\par 

Automatic operation via MCP is a straightforward and efficient approach that should be considered first. However, the MCP server implementation tends to become bloated as the number of operable functions (called MCP tools) increases. Since the list of MCP tools includes metadata such as usage and argument descriptions, adding more supported operations consumes additional LLM context. Moreover, an MCP tool must be implemented for each function, making it labor-intensive to cover the wide range of features available in DCC tools. Therefore, 3Dify also provides an alternative approach.
\par 

\subsubsection{Using CUA}
\label{sec:cua}
In addition to the MCP-based approach, 3Dify also provides automatic operation functionality using the Computer-Using Agent (CUA)\footnote{https://openai.com/index/computer-using-agent/}. CUA is a method by which LLMs directly operate graphical user interfaces (GUIs) through screenshots. 3Dify adopts an LLM specialized for CUA, called UI-TARS~\cite{qin2025uitarspioneeringautomatedgui}, a model fine-tuned from the Qwen-VL series of visual–language models for GUI screen operations. We integrated it into our custom Dify plugin using the SDK.
\par 

Compared with the MCP-based approach, the use of CUA carries a higher risk of misoperation. To mitigate this risk, 3Dify dynamically sets an upper limit on the number of attempts made by the Manager LLM to invoke UI-TARS, taking into account the complexity of the operations assigned to it.
\par 

\subsubsection{Manual Operation of DCC Tools}
For functions that cannot be controlled through either MCP or CUA, 3Dify requests manual operation from the user. Although not yet implemented, conceptually, if collaborative screen interaction is preferable, it may recommend using real-time screen-sharing and voice-dialogue services such as \textit{Share Your Screen}, provided by Google AI Studio\footnote{https://aistudio.google.com/}, which leverages Gemini's multimodal capabilities.
\par 

\begin{figure}[t]
\centering
\begin{lstlisting}%[language=Python]
%\begin{lstlisting}[language=Python, caption = Part of the implemented DCC MCP server template, label = lst:1]

from mcp.server.fastmcp import FastMCP
# Launch MCP server with a name
mcp = FastMCP("3Dify-MCP-Server")

# Execute LLM-generated command on DCC's default console
@mcp.tool()
def run_cmd_on_default_console(cmd: str):
    """Execute command on DCC's default console"""
    # DCC tool-specific implementation

# Get list of configured shortcut keys
@mcp.resource("shortcut://keys")
def get_shortcut_keys() -> dict:
    """Get activated shortcut key list"""
    # DCC tool-specific implementation
    return json 
\end{lstlisting}
\caption{Part of the implemented DCC MCP server template}
\label{fig:mcp_template}
\end{figure}

\begin{figure}[t]
    \centering
    \includegraphics[width=\hsize]{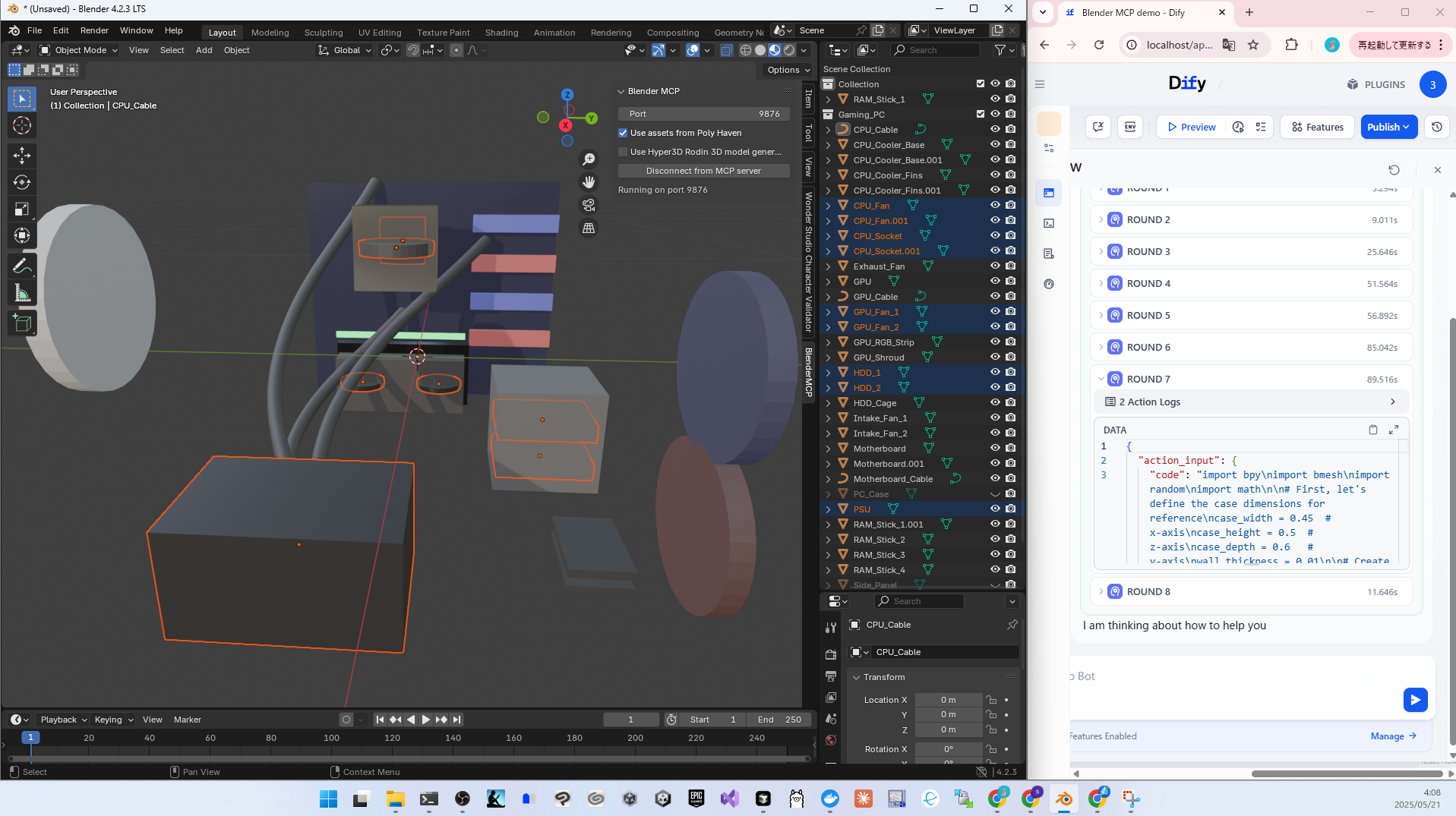}
    \caption{Desktop PC 3D model generated with a single instruction (the authors have hidden the PC case to show the interior)}
    \label{fig:insidePC}
\end{figure}

\begin{figure}[t]
    \centering
    \includegraphics[width=\hsize]{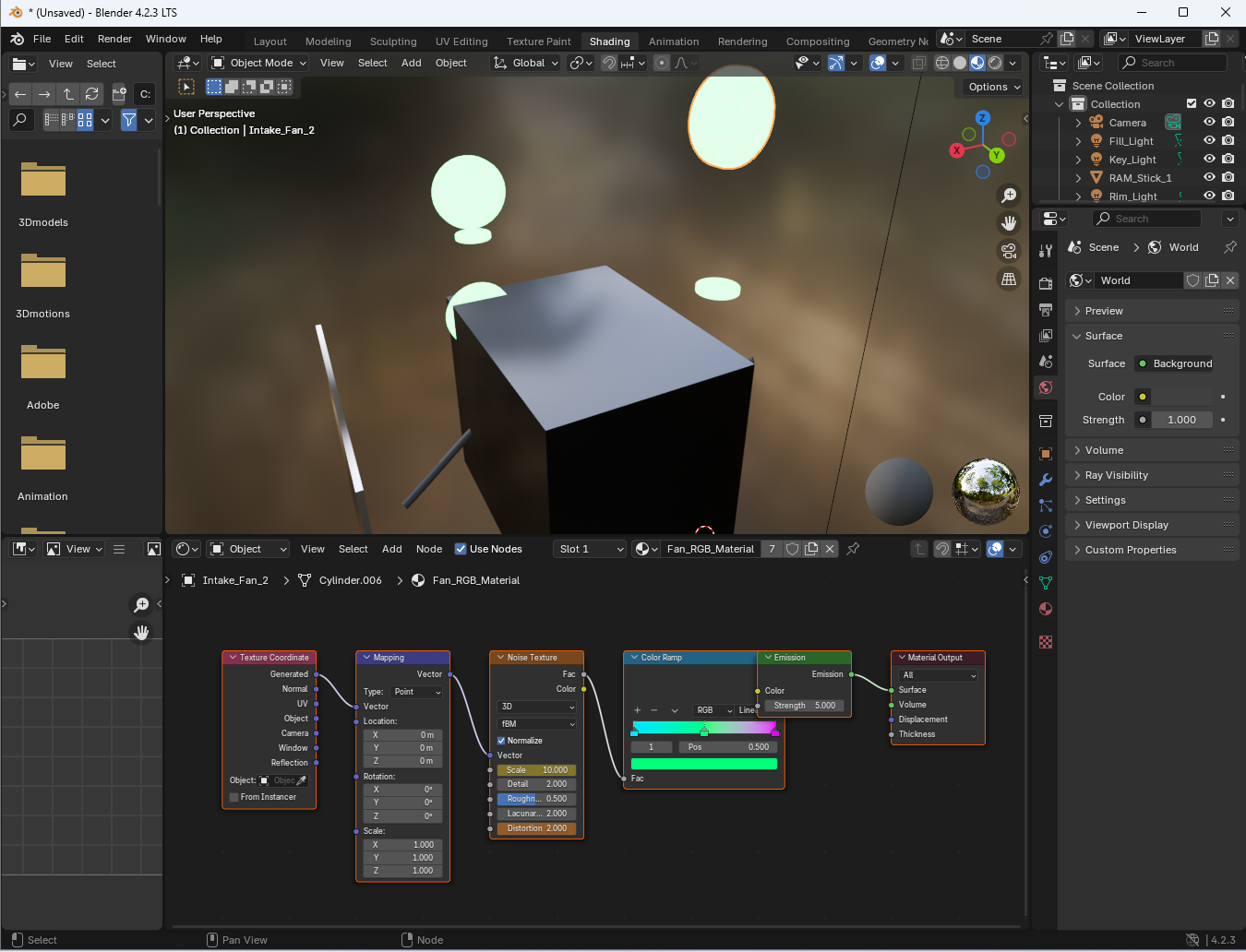}
    \caption{3D model after giving multiple additional instructions}
    \label{fig:outsidePC}
\end{figure}

\section{Demonstration}
\label{sec:demo}
This section demonstrates one of the features of 3Dify, the automatic operation of DCC tools via MCP. The experimental environment is as follows:
\begin{itemize}
    \item Windows 11
    \item Blender 4.2 (LTS)
    \item Dify: v1.3
    \item MCP client: Custom Dify plugin v0.0.4\footnote{https://github.com/3dify-project/dify-mcp-client}
    \item MCP server: ahujasid/blender-mcp\footnote{The latest version as of May 21 of 2025 was used, but the screenshot function was not implemented}
\end{itemize}

Fig.~\ref{fig:insidePC} shows the result of generating a 3D-CG model of a desktop PC from the prompt below.
\begin{lstlisting}%[language=Python]

Create a desktop gaming PC model 
with side panel removed,
keeping all internal components fully visible.

\end{lstlisting}
To make the generated output easier to understand, the PC case was manually removed when taking the screenshot. It should be noted, however, that apart from this adjustment, we did not directly operate Blender at all—the image was generated solely through natural language processing.
\par 

From the LLM's log output, we observed that a large proportion of the calls among the available MCP tools were to the “code” tool. This tool executes Blender operation scripts written in Python, which are generated by the Manager LLM. This is likely because open-source Blender itself provides a well-developed auto-operation API.
\par 

Fig.~\ref{fig:outsidePC} shows the 3D model after several additional instructions were issued. The instruction to make the case fans glow was successful. However, the instruction to move the entire PC upward did not function correctly -- some parts protruded from the case. We infer that while the LLM retained detailed information about object shapes and coordinates during the initial PC part generation, it became difficult to maintain spatial coherence in the subsequent stages. As additional tasks accumulated, the LLM may have struggled to fully track the positional relationships among dozens of objects. We expect this issue to improve with longer LLM context lengths. Another promising direction for improvement is to employ CUA. We emphasize again that, in this demonstration, only MCP was used, and the LLM did not incorporate any visual information in the 3D-CG generation process.
\par 

\section{Conclusion}
\label{sec:conclusion}
In this paper, we proposed 3Dify, a procedural 3D-CG generation framework utilizing Large Language Models (LLMs). 3Dify provides the following key features:
(1) An implementation based on Dify, an open-source AI application development platform, which facilitates the integration of various state-of-the-art LLMs and the orchestration of related technologies such as RAG.
(2) Automated operation of DCC tools via MCP and CUA, enabling users to create 3D-CG using only natural language instructions without manually operating DCC software.
(3) Utilization of RAG to reference DCC resources, thereby improving 3D-CG generation performance.
(4) An interactive feedback loop that refines generated images to better match user preferences.
(5) Support for local LLMs, reducing API costs by leveraging users’ computational resources and enabling the use of custom models.
(6) Automation of DCC tool operations through CUA, allowing access to functions beyond 3D-CG creation and achieving extensibility for applications in broader domains.
Through these features, 3Dify achieves efficient and flexible image generation capabilities not found in previous procedural 3D-CG generation tools. Part of 3Dify has been released as open-source software on GitHub\footnote{https://github.com/3dify-project/}.
\par 

\section*{Acknowledgment}
This research was supported by the Joint Usage/Research Center for Interdisciplinary Large-scale Information Infrastructures (JHPCN) and the High Performance Computing Infrastructure (HPCI) (Project ID: jh250015). This work was also supported by JSPS KAKENHI Grant Numbers JP23K11126, JP24K02945.

\bibliographystyle{IEEEtran}
\bibliography{bibtex}

\end{document}